\documentclass[prl,aps,twocolumn,preprintnumbers,showpacs,superscriptaddress]{revtex4}
\usepackage{latexsym,epsfig,amssymb,amsfonts,amsmath,graphicx,color,bm,times,hyperref,amstext,epstopdf,braket}

\begin{document}

\title{Non-universality of Entanglement Convertibility}

\author{Helena Bragan\c{c}a}
\affiliation{Departamento de F\'isica, Universidade Federal de Minas Gerais, 
Belo Horizonte, MG, Brazil}
\author{Eduardo Mascarenhas}
\affiliation{Departamento de F\'isica, Universidade Federal de Minas Gerais, 
Belo Horizonte, MG, Brazil}
\author{G. I. Luiz}
\affiliation{Departamento de F\'isica, Universidade Federal de Minas Gerais, 
Belo Horizonte, MG, Brazil}
\author{C. Duarte}
\affiliation{Departamento de Matem\'atica, Universidade Federal de Minas Gerais, 
Belo Horizonte, MG, Brazil}
\author{R. G. Pereira}
\affiliation{Instituto de F\'isica de S\~ao Carlos, Universidade de S\~ao Paulo, 
C.P. 369, S\~ao Carlos, SP, 13560-970, Brazil}
\author{M. F. Santos}
\affiliation{Departamento de F\'isica, Universidade Federal de Minas Gerais, 
Belo Horizonte, MG, Brazil}
\author{M. C. O. Aguiar}
\affiliation{Departamento de F\'isica, Universidade Federal de Minas Gerais, 
Belo Horizonte, MG, Brazil}


\date{\today}

\begin{abstract}
Recently, it has been suggested that operational properties connected to quantum computation can be alternative indicators of quantum phase transitions.
In this work we systematically study these operational properties in 1D systems that present 
phase transitions of different orders. For this purpose, we evaluate the local 
convertibility between bipartite ground states. 
Our results suggest that the operational properties, related to 
 non-analyticities of the entanglement spectrum, are good detectors of explicit symmetries of the model, but not necessarily of phase transitions. We also show that thermodynamically equivalent phases, such as Luttinger liquids, may display different convertibility properties depending on 
the underlying microscopic model. 
\end{abstract}

\pacs{03.67.Mn, 64.70.Tg, 75.10.Pq}

\maketitle

Low order phase transitions are directly related to singularities in the derivatives of the 
free energy, with such singularities marking the boundary between two macroscopically 
distinguishable phases. These transitions are usually described by Landau's paradigm of 
symmetry breaking and as such are detectable by local order parameters~\cite{Sachdev}. 
The singular behavior not only manifests itself in thermodynamical properties, 
but also has dynamical consequences, for instance the critical slowing down of adiabatic time 
evolution~\cite{Anatoli,Monta}. 
On the other hand, the existence of phase transitions that do not correspond to symmetry 
breaking is well established~\cite{Wen}. A simple example is provided by the infinite-order 
Berezinskii-Kosterlitz-Thouless (BKT) transition~\cite{BKT} realized in several 
two-dimensional systems at finite temperature, a recent example being cold atomic 
gases~\cite{2DBosegas}. The BKT universality class also turns up in quantum phase transitions 
in one-dimensional (1D) systems such as spin chains~\cite{giamarchi} and Bose-Hubbard chains 
realizable in optical lattices~\cite{Lewenstein}. The existence of such transitions and the 
growing interest in topological phases in 
general~\cite{AliosciaTop,Alioscia,Bonderson,Pollmann2012,Zhao,Zheng}, which cannot be 
described by local order parameters, have motivated 
alternative approaches to address quantum phase transitions~\cite{Gu,Norbert,SymmFid}.

Quantum information theory has provided novel perspectives in this context based, for example,  
on the analysis of the intrinsic correlations (entanglement entropy, entanglement spectrum) of the quantum states of a given system~\cite{EntEnt,Rev,ground_state,Nos,Mund,HaldaneSpec,Pollmann2010, Calabrese2008,Thomale,GGB,Vincenzo,Chiara,Lepori,Giampaolo}.
More recently, a few studies have analyzed the so-called local convertibility of quantum states~\cite{Marcelo,vedral,CuiCao,Alioscia}, which introduces an operational view related to quantum computation. This quantity is completely characterized by functions of the entanglement spectrum, either through majorization relations or the R\`enyi entropies~\cite{Nielsen,Jonathan,Turgut,Sanders}.
Remarkably, it has been shown that several phase transitions coincide with changes in the local convertibility~\cite{Marcelo,vedral,CuiCao}. 

In this Letter we analyze operational aspects of quantum systems using the density matrix 
renormalization group (DMRG) technique. More concretely, we investigate the behavior of the 
local convertibility across several quantum phase transitions in one-dimensional spin-1/2 
and spin-1 XXZ Hamiltonians. Our results show that changes of the local convertibility 
typically correspond to points of symmetric Hamiltonians, which may not coincide 
with quantum phase transitions. 
We also give an example of an infinite-order (BKT) transition that is detached from \textit{any} 
pre-existing symmetries, corresponding to \textit{no} changes in the convertibility profile.
Thus, according to our results the operational approach based 
on convertibility is a good detector of explicit symmetries rather than criticality. 
We clarify that the quantity we consider - the local convertibility - is related to
a protocol of local operations and our main conclusion is that this quantity
indicates explicit symmetries of the Hamiltonians, whose generators are local 
operators (i.e. operators that are sums but never products of single-site operators).
However, hidden symmetries, whose generators are non local operators, are not
directly detected by the convertibility, but through other properties of
the entanglement spectrum.

\textit{Local Convertibility --} Consider a quantum system described by a Hamiltonian $H_{\lambda}$, where $\lambda$ is some tunable parameter.
The system is partitioned into two subsystems, which are distributed to parties $A$ and $B$. The two parties are given the task of converting the ground state $\ket{\Psi_0^{(\lambda)}}_{AB}$ of the initial Hamiltonian into $\ket{\Psi_0^{(\lambda + \epsilon)}}_{AB}$, the ground state of $H_{\lambda +\epsilon}$, by performing only local operations and classical communication (LOCC) on their respective subsystems. The general protocol allows for the use of an extra resource, namely a shared ancillary entangled state called catalyst $\ket{C}$, with which they can freely operate, provided that $\ket{C}$ is left undisturbed in the end of the process, i.e. $\ket{\Psi_0^{(\lambda)}}|C\rangle\rightarrow\ket{\Psi_0^{(\lambda + \epsilon)}}\ket{C}$. This protocol is called local catalytic conversion~\cite{Jonathan,Turgut,Sanders}.  

The necessary and sufficient condition for catalytic conversion relies on the set of R\`enyi entropies, 
\begin{equation}
 S_{\alpha}(\lambda)=\frac{1}{1-\alpha}\log 
\mbox{ Tr}[\rho_{A}^{\alpha}(\lambda)] 
=\frac{\log \sum_i [\xi_{i}(\lambda)]^{\alpha}}{1-\alpha},
\end{equation}
where $\rho_{A}(\lambda)$ is the reduced density matrix of subsystem $A$ and $\xi_{i}(\lambda)$ are its eigenvalues which constitute the entanglement spectrum (ES) of the bipartition \cite{HaldaneSpec}.
Note that $S_0$ is the logarithm of the rank of the state, i.e., the number of nonzero eigenvalues of the reduced density matrix, and $S_{\alpha \rightarrow 1}$ 
is the von Neumann entropy or entanglement entropy (EE), while $S_{\alpha \rightarrow \infty}$ is the logarithm of the largest eigenvalue. 
The condition for conversion is 
$S_{\alpha}(\lambda) \geq S_{\alpha}(\lambda + \epsilon)$ for all $\alpha$~\cite{Turgut, Sanders}, i.e., no entropies can increase after the conversion. In the $\epsilon \rightarrow 0$ limit, 
this relation can be replaced by the analysis
of the signs of the catalytic susceptibility $\chi(\alpha,\lambda)=\frac{\partial S_{\alpha}(\lambda)}{\partial \lambda}$. 
If $\chi<0$ for all $\alpha$, then conversion is only possible
from $\lambda$ to $\lambda +\epsilon$; if $\chi > 0$, $\forall \alpha$, then conversion is possible
only in the opposite direction. For $\chi=0$ conversion is possible in both directions. This criterium was used in~\cite{Marcelo} to analyze the power of adiabatic quantum computation in different phases of a given $\lambda$-parametrized Hamiltonian.

However, some states allow for local convertibility even in the absence of a catalyst. In this case one can use the criterium of majorization of quantum states, defined as~\cite{Nielsen}
\begin{equation}
M(j)=\frac{\partial }{\partial \lambda}\sum_{i=1}^j \xi_{i}(\lambda) ,
\label{majorization}
\end{equation}
where the entanglement spectrum is sorted in decreasing order. Convertibility from $\ket{\Psi_0^{(\lambda)}}$ to $\ket{\Psi_0^{(\lambda + \epsilon)}}$ in the absence of the catalyst is possible if $M(j) \geq 0$ for all $j$. 

\textit{Anisotropic spin chain models --} We consider the XXZ Hamiltonian \begin{multline}
H=\sum_{l=1}^N  [ S^x_{l} S^x_{l+1} + S^y_{l} S^y_{l+1}
 +\Delta S^z_{l} S^z_{l+1} + D ({S^z_l})^2] ,
\label{XXZ}
\end{multline}
where $\mathbf S^{(x,y,z)}_l$ are the spin operators at the $l$th site, $\Delta$ is the longitudinal nearest-neighbor 
exchange interaction, and $D$ represents uniaxial single-ion anisotropy. In eq.~\ref{XXZ} 
and through out this paper, we assume the transverse nearest-neighbor exchange interaction 
as the unit of energy.
We study both spin-1/2 and spin-1 systems; for the former the single-ion anisotropy term is simply a constant, since $\sigma_z^2=1$ (where $\sigma_z$ is a Pauli matrix). 
We shall identify the control parameter $\lambda$ with either $\Delta$ or $D$, depending on the transition in question. 
For any values of $\Delta$ and $D$, the model has an explicit U(1) rotational symmetry in the $xy$ plane and a $Z_2$ spin inversion symmetry along the $z$ axis. 

The spin-1/2 model is integrable and exactly solvable by Bethe ansatz ~\cite{giamarchi}.
The system has two gapped phases: 
a ferromagnetic phase for $\Delta < -1$ and an antiferromagnetic (AFM) N\'eel phase for $\Delta > 1$, separated by a gapless Luttinger liquid (LL) XY phase for $-1<\Delta \leq 1$. The phase transition between the 
ferromagnetic and LL phases is of first order, while the one between LL and AFM
phases is a BKT transition. In the latter the gap decreases exponentially as $\Delta\to 1^+$, which poses a challenge for numerical techniques attempting to detect  the critical point~\cite{Nomura1, GGB}. The model has an explicit SU(2) symmetry at $\Delta=1$ (both in the spin-1/2 and spin-1 cases), that is, there exists a set of four operators $Z=\sum_lS^z_l$ (the total Z-spin), $S^2$ (the total spin), $Q=\sum_l S^-_l=\sum_l S^x_l -i S^y_l$ (a lowering operator) and $Q^{\dagger}$
which commute with the Hamiltonian at $\Delta=1$ and are constructed as sums over local
operators.

The spin-1 model is not integrable, but its ground state phase diagram is known \cite{dennijs,diagrama}. Let us first focus on the $D=0$ line. In this case one finds a ferromagnetic phase for $\Delta\leq-1$, a LL XY phase for $-1<\Delta\leq0$, a Haldane phase for $0<\Delta\lesssim 1.18$ and a N\'eel phase for $\Delta\gtrsim1.18$ \cite{Nomura3, YHSu}. The gapped Haldane phase is characterized by a nonlocal string order parameter that breaks a hidden $Z_2\times Z_2$ symmetry \cite{Kennedy} and is an example of a symmetry protected topological phase \cite{Pollmann2012}. Most interestingly, the transition between the XY and the Haldane phase is of BKT type and is known to occur exactly at $\Delta=0$ due to a hidden SU(2) symmetry generated by nonlocal operators \cite{Nomura4}. 
To be more specific, in this case the $Q$ operators that define the algebra satisfy 
$Q^+ \propto\sum_j(S_j^+)^2e^{i \pi\sum_{l<j}S^z_l}$, $Q^- \propto\sum_j(S_j^-)^2e^{i \pi\sum_{l<j}S^z_l}$ 
and $Q^z \propto [Q^+, Q^-]$, that is, they involve sums over nonlocal operators, in contrast to the case of
the explicit SU(2) symmetry present at $\Delta=1$.

Switching on the single-ion anisotropy in the spin-1 model gives rise to a BKT transition line in the phase diagram \cite{Nomura3}. While the transition between XY and Haldane remains pinned at $\Delta=0$ due to the hidden SU(2) symmetry, there appears a BKT transition between the XY phase and a so-called large-$D$ phase, favoured by strong easy-plane anisotropy ($D>0$). The gapped large-$D$ phase is topologically trivial as the nondegenerate ground state is adiabatically connected to the product state with $S_l^z=0$ for all spins. The transition from XY to large $D$ is completely dissociated from high symmetry points in the lattice model. It is important to distinguish between the exact SU(2) symmetry at $\Delta=0$ \cite{Nomura4} and the SU(2) symmetry that arises in the renormalization group analysis of the sine-Gordon model, which is the effective field theory for the BKT transition \cite{giamarchischultz,Nomura2}. This emergent symmetry is a genuine signature of a quantum phase transition since it 
becomes asymptotically exact in the thermodynamic limit.

\begin{figure}
\includegraphics[width=\linewidth]{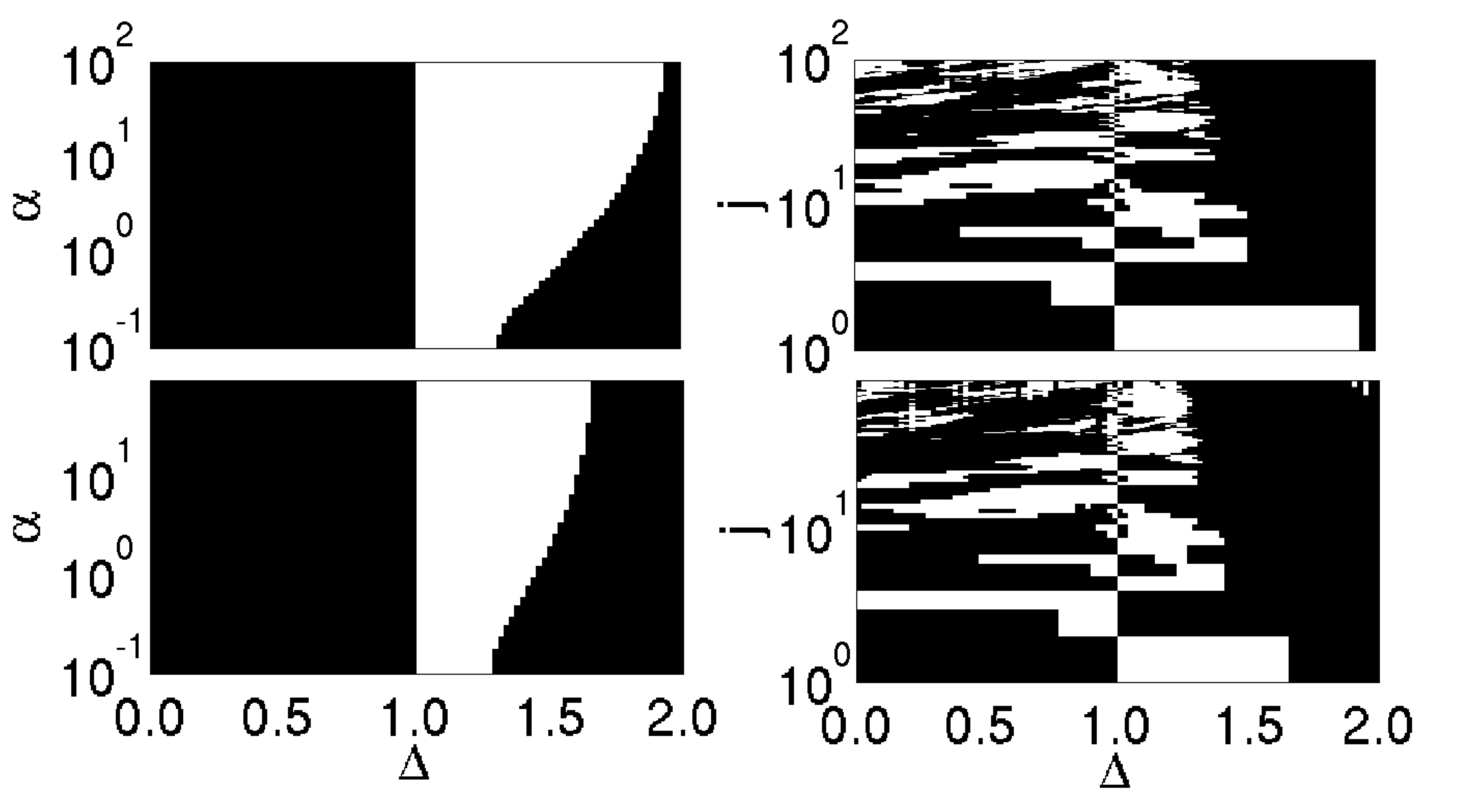}
\caption{The sign of the catalytic susceptibility $\chi $ for different R\`enyi entropies (left panels) and majorization $M$ (right panels) 
as a function of $\Delta$ for a spin-1/2 chain with $N=112$ (top) and $N=212$ (bottom), where $N$ is the size of the chain.
The black region indicates $\chi<0$ (or $M<0$)
and the white region corresponds to $\chi>0$ (or $M>0$). The
data was generated for a symmetrical bipartite system $A=B=L/2$, with a DMRG truncation error below $10^{-9}$.} 
\label{conv100}
\end{figure}

\textit{Operational properties --} For both spin-1/2 and spin-1 models, the ground state is always one of the two degenerate fully polarized states for any $\Delta\leq-1$. Thus, in the ferromagnetic phase $\chi=0$ and convertibility is possible between any two points (in both directions). 

The other phases have nontrivial convertibility behavior. First consider the spin-1/2 model. 
In the left panels of Fig. \ref{conv100} we present the sign of the catalytic susceptibility obtained using DMRG for open chains with $N= 112$ sites (top panel) and $N=212$ (bottom panel). 
Differently from the ferromagnetic phase, for $-1 < \Delta < 1$ (which corresponds to the critical LL phase in the thermodynamic limit), $\chi<0$ for all $\alpha$, which indicates unidirectional convertibility.
In addition, the convertibility changes direction at the isotropic point $\Delta=1$.  
The convertibility is then lost (i.e., the sign of $\chi$ depends on $\alpha$) as $\Delta$ increases and it is recovered (unidirectionally) for larger $\Delta$ as
the ground state approaches the classical N\'eel state. 
The convertibility is sensitive to the chain length: the larger the system, the smaller the value of $\Delta$ for which it presents the 
convertibility  characteristic of  strong AFM behavior. 

Results for the majorization analysis are shown in the right panels of Fig. \ref{conv100} for both $N=112$ and $N=212$,
from which we can see that the use of a catalyst in the conversion is dispensable   only in the strong AFM regime. 

We stress that, independently of the chain length, the catalytic convertibility changes direction
exactly at the Heisenberg point $\Delta=1$ (see Fig. \ref{conv100}), which coincides with the 
SU(2) symmetry of the model. The absence of finite size effects at this point, even for small 
chains with $N\sim 10$ sites, suggests that the convertibility is detecting the SU(2) symmetry
(expected to be present for chains of \textit{any} size), rather than the phase transition, 
whose precursors should be apparent only for large systems. It is also interesting 
that the majorization relations (also in Fig.~\ref{conv100}) display a local mirror-like 
symmetry around $\Delta=1$, which reinforces that these quantities detect the SU(2) symmetry.

\begin{figure}
\includegraphics[width=\linewidth]{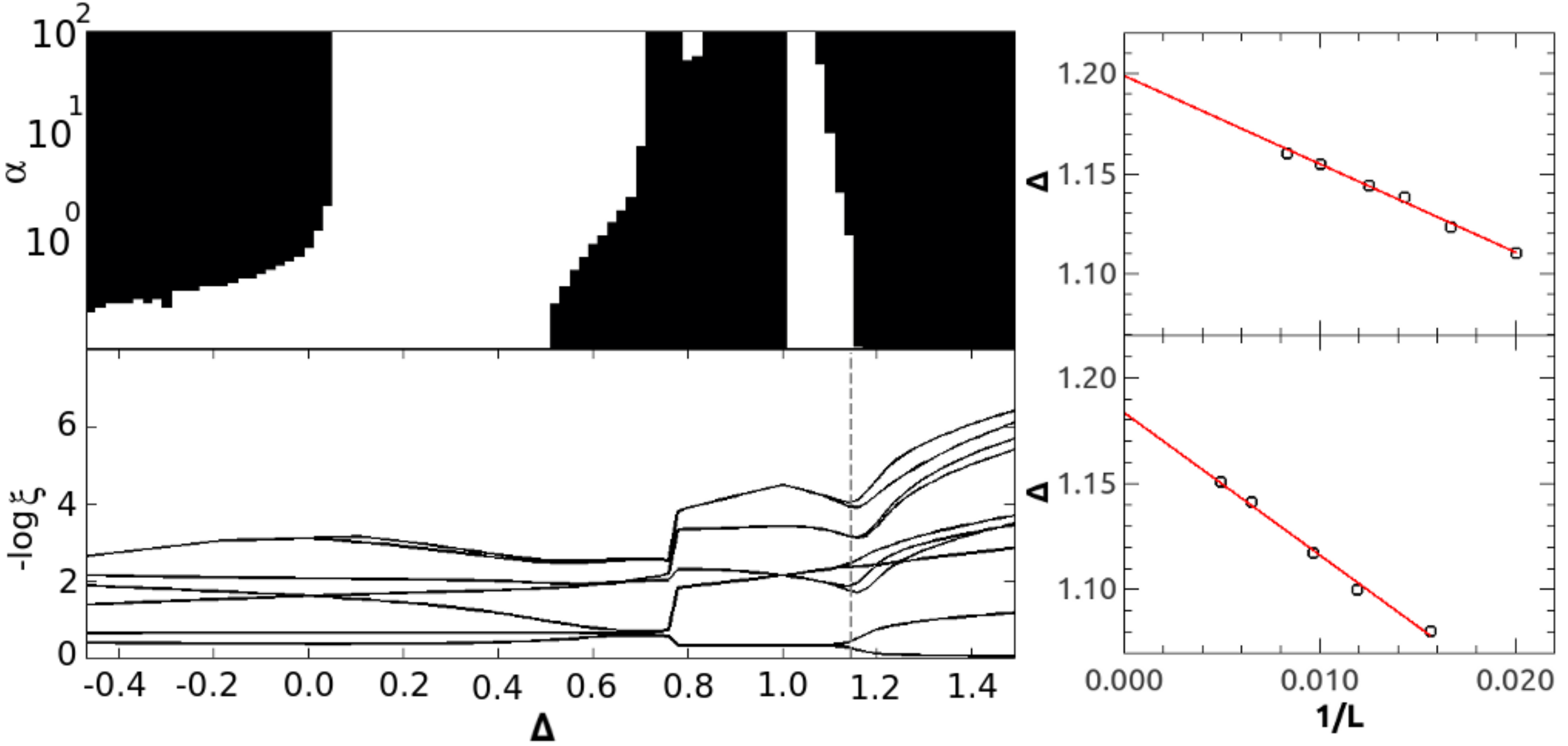}
\caption{Sign of the catalytic susceptibility for different R\`enyi entropies (left-top) and entanglement spectrum (left-bottom) of a symmetrical bipartite $A=B=L/2$ spin-1 system as a function of $\Delta$, for  $D=0$ and $N=106$. 
The right panels show finite size scaling procedures corresponding to the point where 
the unidirectional convertibility is recovered (top) and the level degeneracies are 
lifted (bottom) when going from the Haldane phase to the N\'eel phase.} 
\label{Convertibility100S1m100}
\end{figure}

Let us now discuss the $D=0$ spin-1 model.
Our results for the convertibility properties in this case are presented in the top panel of Fig. \ref{Convertibility100S1m100}. 
Interestingly, the LL phase ($-1<\Delta <0$) is not locally convertible, since the sign of $\chi$ depends on $\alpha$,
unlike the LL  phase in the spin-1/2 model, which is convertible. This is a  remarkable fact: even though the LL phases of spin-1/2 and spin-1 models share the same low-energy physics, they exhibit   different operational behavior. Nevertheless, we note  that the catalytic susceptibilities corresponding to large values of $\alpha$, which are dominated by the largest eigenvalues of the density matrix, have the same sign ($\chi<0$) for both spin-1/2 and spin-1 LL phases. This seems consistent with the general expectation that universal information can be extracted from the low-lying levels in the ES \cite{HaldaneSpec,Calabrese2008}. In contrast, the low-$\alpha$ (i.e., ``high temperature'' \cite{HaldaneSpec}) susceptibilities may depend on details of the microscopic model.

Now, the transition from the LL to the Haldane phase (known to occur at $\Delta=0$ in the
thermodynamic limit) does not coincide with a change in the convertibility (see 
Fig. \ref{Convertibility100S1m100}). At this point the Hamiltonian presents the hidden SU(2) 
symmetry (related to \textit{nonlocal} operators) 
discussed previously in this paper. This \textit{hidden} symmetry is not detected 
by the \textit{local} convertibility. Nonetheless, it is important to stress that the 
EE ($S_{\alpha\rightarrow1}$) does present a derivative that changes sign exactly 
at the symmetry point, which also coincides with a level crossing in the ES (bottom of the
figure). 
For $\alpha<1$, the susceptibilities change sign for $\Delta<0$, while for $\alpha>1$ the sign 
changes happen for $\Delta>0$. Thus, even though the convertibility is blind to this 
hidden symmetry, it is still encoded in the ES.

Figure \ref{Convertibility100S1m100} (top) also shows changes in the convertibility profile within the Haldane phase. 
The unidirectional convertibility is established approximately once the phase is reached ($0\lesssim\Delta \lesssim 0.5$), but it is then lost
for $0.5 \lesssim \Delta \lesssim 0.7$. As $\Delta$ increases even further, still in the Haldane phase, the convertibility is recovered in the opposite direction. 
More interestingly, $\chi$ changes sign for all $\alpha$ exactly at the SU(2) symmetry point $\Delta=1$. 
We emphasize that all these alterations happen inside the same phase, leading us to conclude that changes in the local convertibility do not 
necessarily correspond to phase transitions. Furthermore, there seems to be a direct relation between local convertibility and symmetries related to local operators.

The convertibility profile is a function of the ES. The $10$ largest eigenvalues of the reduced density matrix are presented in the bottom panel of Fig. \ref{Convertibility100S1m100}.
As we can see, the most abrupt changes in the convertibility properties occur in the vicinity of points of either level crossing (e.g. near $\Delta=1$) or degeneracy breaking of the ES (e.g. near $\Delta\approx   1.18$), that is, are related to non-analyticities of the ES. 

It is interesting to understand why there are level crossings at $\Delta = 0$
and $\Delta = 1$, which correspond to the SU(2) symmetry points. The system ground state is an eigenvector of the symmetry operators and the partition we consider
does not break the symmetry. This means that the left and right eigenvectors of the Schmidt decomposition 
can be identified by the symmetry quantum numbers. For the SU(2) symmetry, energy-eigenvectors 
corresponding to different eigenvalues of $S_z$ can be degenerated if they are connected 
by the $S^+$ or $S^-$ operators defined at each partition. This gives rise to the level crossings we observe in Fig. \ref{Convertibility100S1m100}. As these crossings are associated with the symmetry, 
which is present for systems of any size, we do not expect these crossings to shift with 
the system size, as indeed observed in our numerical results (not shown). We add
that the connection 
between symmetry and degeneracy in the EE has been analyzed in detail 
in other cases \cite{crossing}.

A second order (Ising type) phase transition from the Haldane to the 
N\'eel phase happens for $\Delta \approx 1.18$ (in the thermodynamic limit) 
\cite{Nomura3, YHSu}. As can be seen in the left-top panel of 
Fig. \ref{Convertibility100S1m100}, 
around this value of $\Delta$ we observe a change in the convertibility sign. Contrary 
to the cases we analyze above, here the convertibility profile \textit{is} sensitive 
to finite 
size effects. Note that this phase transition falls under the standard symmetry breaking 
paradigm and, if the convertibility is able to detect the transition, it should depend
on the system size, as it indeed does. 

In previous literature \cite{Marcelo, CuiCao}, changes in the convertibility profile
are associated to this symmetry breaking type of transition, but only 
small systems are analyzed (up to 18 sites).  
Here we consider larger systems and perform a finite-size scaling analysis, which can 
be seen at the right-top panel of Fig. \ref{Convertibility100S1m100}. In the 
thermodynamic limit, the critical $\Delta$ obtained from our scaling procedure is 
$\Delta \approx 1.20$, which slightly deviates from the known value of $\Delta\approx1.18$. 
Indeed, we do not expect a very good estimation of the critical $\Delta$ from 
the convertibility, since the changes in its profile correspond to R\`enyi entropies
of small-$\alpha$ that strongly depend on the smallest reduced density matrix 
eigenvalues, which intrinsically have less accuracy in the DMRG calculation.

The largest eigenvalues of the reduced density matrix are numerically more precise 
and indeed the 10 largest values of them can be used to better determine where this 
Ising type of transition happens. The change in the convertibility analyzed above
corresponds to a level splitting in the ES - see the dashed line in the left-bottom 
panel of the figure. This degeneracy breaking can be associated with the Haldane-N\'eel 
phase transition since for open chains the higher degeneracy of the Haldane phase is 
attributed to the spontaneous breaking of the hidden $Z_2\times Z_2$ symmetry 
\cite{Pollmann2010}. The value of $\Delta$ where this degeneracy is lift also 
\textit{shifts} with the system size and a finite size scaling allows us to 
obtain the corresponding value in the thermodynamic limit. As can be seen in the 
right-bottom panel of Fig. \ref{Convertibility100S1m100}, our analysis yields 
$\Delta \approx 1.18$, in good agreement with the known critical value for the 
Haldane-N\'eel phase transition~\cite{Nomura3, YHSu}. 

We are thus able to correctly identify the transition 
point by analyzing the ES, which is directly related to the convertibility profile. 
This result and the previous literature suggest that 
the convertibility is a detector of criticality only in the case of transitions associated 
with symmetry breaking. In most of the cases, however, it is a detector of explicit 
symmetries of the Hamiltonian.

Regarding the majorization relations, our results for the spin-1 system
indicate that convertibility without a catalyst is possible only in the strong AFM 
(large $\Delta$) regime, similarly to the N\'eel phase of the spin-1/2 model.

\begin{figure}
\includegraphics[width=\linewidth]{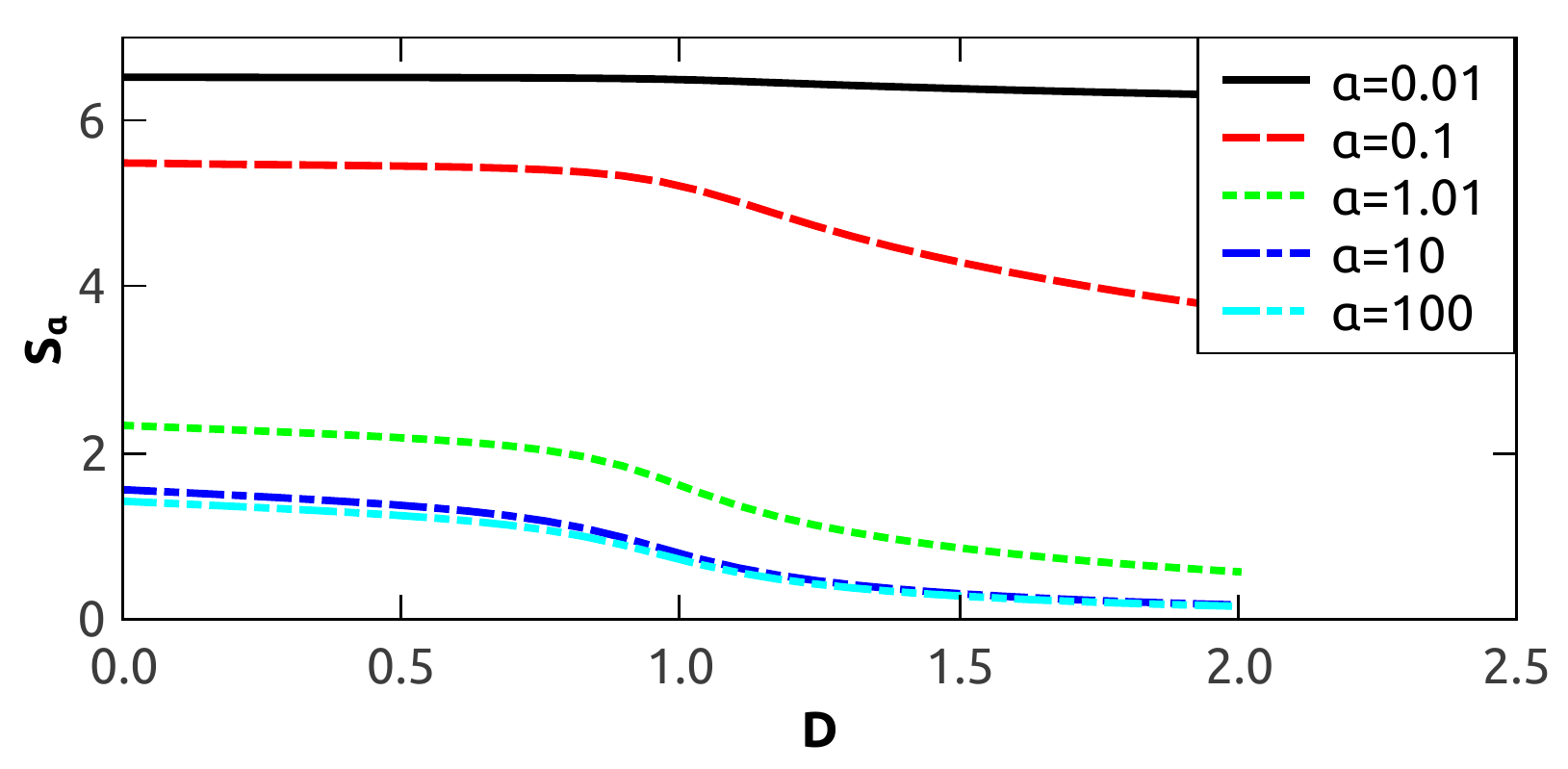}
\caption{(color online). R\`enyi entropies $S_{\alpha}$ ($\alpha=0.01;0.1;1.01;10$ and $100$) for a $N=206$, spin-1 chain with fixed $\Delta=-0.5$
and $A=B=L/2$, as a function of $D$. 
The entropies are always monotonic, which means that there is no change in the direction of the convertibility.} 
\label{S1_renyi}
\end{figure}

\textit{Single-ion anisotropy -- }To strengthen the conclusion drawn above that the convertibility 
is a detector of symmetries instead of phases transitions, we study the spin-1 XXZ chain with 
single-ion anisotropy. As mentioned above, the phase diagram for $\Delta < 0$ and $D>0$ contains a BKT transition (without symmetry breaking) from a critical XY phase to a gapped large-$D$ phase \cite{diagrama}. This transition does not coincide with any high symmetry points in the Hamiltonian for finite chains.  
In Fig.~\ref{S1_renyi}, we show results for some R\`enyi entropies as a function 
of positive $D$, for $\Delta=-0.5$. The main point is that all entropies 
decrease monotonically with $D$, leading to a uniform convertibility profile 
($\chi<0$ for all $\alpha$ and all $D$). The BKT transition is expected to happen at $D \approx 0.8$ for $\Delta=-0.5$~\cite{diagrama}, but there is no sign of it in the convertibility 
profile. We stress that this is the first example (to our knowledge) of a phase transition around which there is absolutely no change in the convertibility.

\textit{Critical Entanglement entropy -- } 
The phase transition analyzed in the last section constitutes an important 
example in the context of our work, since it is not accompanied by a change in the 
convertibility. This infinite order phase transition which does not have a pre-existing 
SU(2) symmetry has, however, a small amount of work dedicated to it. Here we show that 
the EE $S_{\alpha\to1}$ can be used to detect it through a simple finite size analysis.

\begin{figure}[h]
\includegraphics[width=\linewidth]{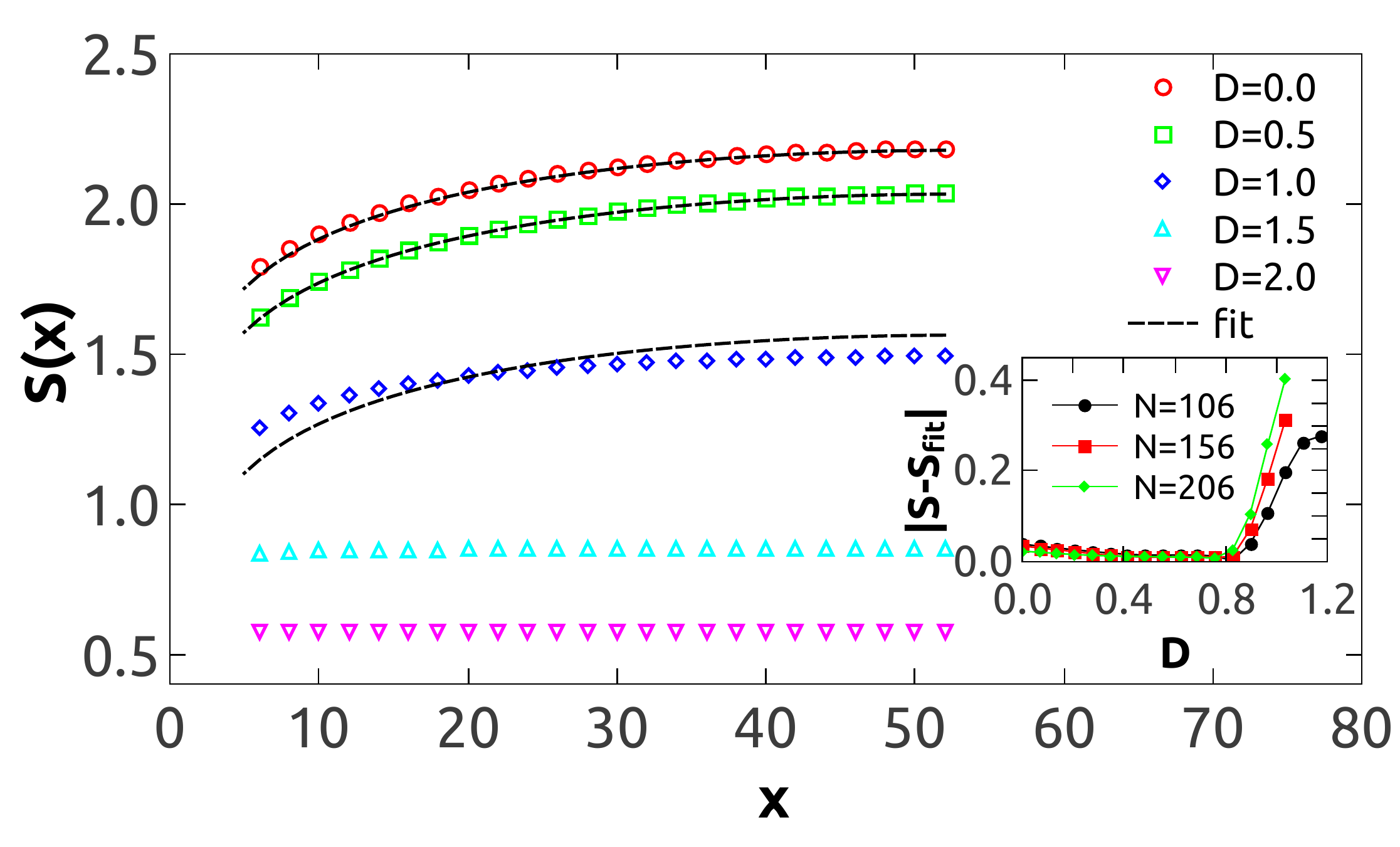}
\caption{(color online). Entanglement entropy as a function of the partition size 
$x$ for $N=106$, $\Delta = -0.5$ and different values of the single-ion 
anisotropy $D$. The fittings of the numerical data to eq. \ref{OBC} are also shown.
The inset presents the fitting accuracy for systems of three sizes ($N=106, 156, 206$), from which we find the BKT critical point 
at $D \approx 0.8$.} 
\label{Fit_entropy}
\end{figure}

In Fig.~\ref{Fit_entropy} we show the EE, $S(x)$, as a function of the partition size 
$x$ for $N=106$, $\Delta = -0.5$ and different values of the single-ion 
anisotropy $D$. It is clear that the behavior of $S(x)$ qualitatively changes with 
$D$: it increases logarithmically with $x$ for small $D$, but saturates 
for large $D$ values, indicating a phase transition.  

In fact, the EE is specially useful since it has been shown to exhibit universal scaling in the LL phase, which is described by a conformal field theory (CFT) with central charge $c=1$. Using CFT, Calabrese and Cardy~\cite{Calabrese2004} showed that the EE of a finite system with open boundary conditions in the regime $ x,N \gg 1$ is given by 
\begin{equation}
S(x,N)=\frac{c}{6}\mbox{log}\left[\frac{2N}{\pi}\mbox{sin}\left(\frac{\pi x}{N} \right) \right]+ s',
\label{OBC}
\end{equation}
where $s'$ is a non universal constant.

We can find the BKT critical point by fitting the numerical data to Eq.~\ref{OBC}, 
leaving $c$ and $s'$ as free parameters, as shown in Fig.~\ref{Fit_entropy}.
For small $D$, the numerical data are well fitted by Eq.~(\ref{OBC}), 
indicating that the system is in the critical phase. As $D$ increases, however, the 
behavior of $S(x)$ starts to deviate from the CFT prediction. 
The fitting accuracy (inset of the figure) clearly shows that Eq.~(\ref{OBC}) correctly
describes the EE 
for $D \lesssim 0.8$, leading us to estimate that the BKT critical point is at 
$D = 0.82 \pm 0.07$, in accordance with previous results~\cite{diagrama}.    
Inside the critical region, the values of the free parameters are as expected: 
$s'$ is independent of the chain length and $c \approx 1$.
  
\textit{Conclusions --}The local convertibility of many-body quantum states, an 
important concept in quantum information and computation 
theory, provides a comparison of the computational potential of adiabatic 
and LOCC procedures.
The results presented in this Letter strongly suggest that this operational 
perspective on quantum states is not necessarily related to quantum phase transitions, but reflects properties of the entanglement spectrum which are intimately connected with symmetries of the microscopic model.
In fact, we explicitly showed examples of local convertibility changes
that do not correspond to phase transitions and phase transitions that do not 
correspond to alterations of the local convertibility. Furthermore, different models that fall into the same universality class, such as Luttinger liquids, may exhibit distinct convertibility properties. Hence, the non-universality of the convertibility.

Upon preparation of this manuscript we became aware of~\cite{Chandran}, where 
the entanglement spectrum for a couple of systems is shown to display pseudo 
transitions that do not correspond to physical phase transitions. The authors
then conclude that it may be misleading to use entanglement measurements as 
detectors of quantum phase transitions, in agreement with our conclusions. 
Note that their conclusions are based on a different approach than ours, that is, 
one not related to high symmetry points or operational aspects.    

We thank Francisco C. Alcaraz, Raphael Drumond and Mauro Paternostro for useful discussions and S. Montangero
for extensive and intensive tutorials on DMRG. This work 
was supported by CNPq and CAPES. M.C.O.A. and M.F.S. also acknowledge  
FAPEMIG for financial support.

\end{document}